\documentclass{icrc}

\usepackage{times}
\usepackage{graphicx} % when using Latex and dvips
\begin{document}

\title{LogN--LogS Studies of EGRET Sources}
\author[1]{O. Reimer}
\affil[1,2]{NRC/NASA/GSFC, Code 661, Greenbelt, MD 20771, USA}
\affil[2]{Now at: Institut f\"ur Theoretische Physik, Lehrstuhl
IV: Weltraum- \& Astrophysik, Ruhr-Universit\"at Bochum, D-44780 Bochum,
Germany}
\author[3]{D.J. Thompson}
\affil[3]{NASA/GSFC, Code 661, Greenbelt, MD 20771, USA}

\correspondence{olr@egret.gsfc.nasa.gov}

\firstpage{1}
\pubyear{2001}

% \titleheight{11cm} % uncomment and adjust in case your title block
                     % does not fit into the default and minimum 7.5 cm

\maketitle

\begin{abstract}
A comprehensive investigation of logN--logS distributions of gamma-ray sources 
discovered by EGRET has been performed for subsequent use in population studies. 
Existing models explaining the spatial arrangement of unidentified sources do 
not compare against an observed logN--logS distribution. However, viable population 
models not only have to reproduce the logN--logS distribution for different 
source classes globally, they have to correspond to apparent differences among 
their spatial, spectral and variability characteristics. Furthermore, it needs to 
be understood in which way results from selections among the unidentified sources 
like "persistent" \citep{gre00} or "steady" sources \citep{geh00} are related to 
the overall picture regarding their logN--logS characteristics.  
\end{abstract}

\section{Introduction}

A Log N--log S study of any class of astronomical object, if complete for a set of
selection criteria, is a valid and useful tool for diagnoses of source properties.
The completeness of the EGRET detected gamma-ray sources has been probed by analyzing
gamma-ray excesses down to TS values of 9, i.e. below the detection criterion
for inclusion of gamma-ray sources into the 3EG catalog \citep{har99}. On the
condition that completeness for an investigation of 3EG catalog sources could
be obtained this way, various questions concerning the properties of gamma-ray
sources could be addressed. Spatial arrangements, source identification issues,
source fluxes as well as selections of particular interest among the unidentified 
gamma-ray sources have been obtained and visualized in their logN--logS distribution. 
To date, logN--logS studies are not given in the full context of available
gamma-ray observables : \citet{oez96} investigated high latitude AGN and 
unidentified sources on the basis of the 2EG catalog, just available at this time, 
\citet{mue00} analyzed AGN and FSRQs in respect of their different contribution 
to the extragalactic diffuse gamma-ray background, and \citet{geh00} used the 
subset of "steady" unidentified sources to distinguish between faint mid-latitude 
sources and bright unidentified sources close to the Galactic disk. 
Here we want to study the relevant information concerning gamma-ray point source 
detections by EGRET in terms of logN--logS distributions. Scope of this 
investigation is the latter use in a source population model, where observed 
logN--logS distributions need to be adequately reproduced to reflect the reality 
and correspond to instrumental selection effects and detectability biases in 
population models as well.

\section{Analysis \& Results}

Originating from a source list of 416 gamma-ray excesses with test statistics (TS)
greater than 9 (i.e. 3$\sigma$ detections), source fluxes were determined using 
EGRET data from CGRO observation cycles 1 to 4. Identifications are taken from the 
3EG catalog, which made use of a more strict source detection criterion. 
Hence, 263 gamma-ray point sources from the 3EG were used (neglecting 7 artifacts and a 
solar-flare detection). Individual source identifications beyond the 3EG catalog are 
consistently incorporated.
Fig.1 gives the overall picture using different detection significances for the
complete sample of gamma-ray sources. The different significance levels indicate 
the appearance of limitations in EGRETs source detectability, significantly
flattening the distribution at lower fluxes than $2.5\times 10^{-7}$ photons cm$^{-2}$ s$^{-1}$.
We assume completeness for fluxes above this level within given statistical uncertainties.
Fig.2 shows the 3EG catalog sources. Here we compare the discrepancy between
the usage of average flux levels (P1234) and peak fluxes whenever the peak flux exceeds
an average flux level. The instrumental detectability obviously lowered further to about 
$5\times 10^{-7}$ photons cm$^{-2}$ s$^{-1}$, actually reflecting the way the 3EG catalog
has been compiled using sources matching the detection threshold in either individual 
viewing periods or superpositions on timescales up to four years of CGRO operation, 
which is referenced by "P1234". It can bee seen, that the slope of the distribution above 
fluxes influenced by instrumental biases remains the same within the given statistical uncertainties.
Before splitting the dataset into different categories in respect of their identification, 
a check has been done comparing the AGN identifications as given in the 3EG catalog 
(66 high-significance and 27 lower confidence identifications) and their subsequent 
quantitative evaluation by \citet{mat01}. As important the question of a correct 
identification of a gamma-ray source with an AGN is, individual discrepancies do not 
implicate a significant change in the shape of the logN--logS distribution. This is due
to the fact that all high-confidence AGN identification have been confirmed by \citet{mat01}.
Therefore misidentifications could only occur for low flux sources, simply resulting in a 
reduction in N.
Fig.3 finally compared the identified AGN in average and peak flux, respectively. As also seen in Fig.2,
the difference between both curves is significant, whereas the slope of a best-fitted power law not.
The linear fit to the data down to the instrumental detectability bias appears to be 
consistent with the expected $\rm{S}^{3/2}$ dependence for an isotropic/spatial uniform distribution of 
underlying objects for a Euclidean universe 
A similar discrepancy could be noticed when comparing average and peak fluxes of unidentified 
gamma-ray sources. However, the right flank of the logN--logS distribution is heavily 
influenced from the degree of completeness of gamma-ray source identifications itself, which is 
essentially unknown as in the nature of a source being unidentified. Fig.4 is given here only 
as reference for the following latitudinal selections among unidentified sources.
In Fig.5 high-latitude unidentified sources were selected ($\rm{|b|}> 30^{\rm o}$). Two noticeable
differences compared to Fig.3 can be seen: (1) AGN extend to higher flux levels compared to 
unidentified sources at high-galactic latitudes. (2) The contrast between average and peak flux
representation in a logN--logS diagram between unidentified gamma-ray sources and AGN 
is even more pronounced. This is due to the fact, that sources at high-galactic latitude are
preferential identified by its peak flux, therefore leaving the average flux distribution in
a random shape in respect of its completeness in identification. The similarity
between the peak-flux distribution for unidentified high-latitude sources and AGN confirms 
that AGN are the obvious potential identification for these unidentified gamma-ray sources.
Fig.6 shows unidentified sources at mid-galactic latitudes. Here we have chosen three different
latitudinal selections matching previously studied mid-latitude unidentified sources 
($10^{\rm o} >\rm{|b|}> 30^{\rm o}$ reflecting the threshold between two different detection 
criteria as present in the 3EG catalog at $10^{\rm o}$, $5^{\rm o} >\rm{|b|}> 30^{\rm o}$) to compare
with "steady" unidentified sources \citep{geh00}, and $2.5^{\rm o} >\rm{|b|}> 30^{\rm o}$ 
reflecting a cut for "persistent" unidentified sources as used by \citet{gre00}. In fact, the 
difference in the shape of the distributions between these mid-latitude selections are 
within the statistics of the sample itself, therefore it does not matter which cut has been 
chosen in order to draw conclusion from logN--logS distributions.
In Fig. 7 we compare unidentified sources close to the Galactic Plane in average and peak
flux, respectively. Here one clearly sees the suppression from EGRETs inability to
discriminate sources in the Galactic Plane at a similar level against the dominant galactic 
diffuse background compared to sources outside the Plane. Furthermore, the slope of the logN--logS
distribution appears to be rather different for high source fluxes compared to Fig.4. It remains to
be investigates whether this is due to a different class of objects responsible for this steeper
slope as suggested by \citet{geh00} or it is a result from selection effects due to a 
different level of completeness in source identifications within the Galactic Plane. 
Differences between unidentified sources and samples like "steady" unidentified
sources and "persistent" unidentified sources have been looked at, too. When comparing its 
particular logN--logS distribution, the way these selections are compiled is apparent. They basically
resemble the logN--logS distribution for unidentified gamma-ray sources using average flux
values, sorting out unidentified sources which have been included in the 3EG catalog because
of their peak flux instead of cumulative added source significance (in other words: sources
matching the 3EG catalog criteria with its P1234 average flux). Therefore, a comparison with
these samples is meaningful only considering average flux values in logN--logS distributions.
Finally, Fig.8 compares "steady" unidentified sources in respect of its photon spectral index. 
A separation in $\gamma$= -2.25 has been made and the difference in the
logN--logS distributions is clearly seen. Unidentified sources with hard power law spectrum show
a distinct steeper logN--logS distribution than softer unidentified sources.
However, in this case we also have to consider an apparent selection effect: In regions of
dominant Galactic diffuse emission hard sources are easier detectable compared to soft spectrum
gamma-ray emitters, but only to a flux level significantly higher than achievable at high-galactic
latitudes. Therefore differences in the logN--logS distribution could only be discussed meaningfully 
if they are determined within regions of comparable diffuse background level.
Unfortunately, the EGRET data do not allow statistically significant results at this depths
of investigation. These distinctions could only be made when next-generation gamma-ray instruments 
like GLAST will provide much more source detections and subsequently more source identifications, too. 

% (i) one column figure, will be floated to top of next column
%
% \begin{figure}[t]
% \vspace*{2.0mm} % just in case for shifting the figure slightly down
% \includegraphics[width=8.3cm]{figfile.eps} % .eps for Latex,
%                                            % pdfLatex allows .pdf, .jpg, .png and .tif
% \caption{Figure caption text.}
% \end{figure}

% (ii) two column figure, will be floated to top of next page
%
% \begin{figure*}[t]
% \includegraphics[width=17.0cm]{figfile.eps}
% \caption{Figure caption text.}
% \end{figure*}

% (iii) 1 1/2 column figure with caption on the side, will be floated
% to top of next page
% \begin{figure*}[t]
% \figbox*{}{}{\includegraphics*[width=11.0cm]{figfile.eps}}
% \caption{Figure caption text.}
% \end{figure*}

\begin{acknowledgements}
O.R. acknowledges a National Research Council Associateship Award taken at
NASA Goddard Space Flight Center.
\end{acknowledgements}

\begin{figure}[t] 
\vspace*{2.0mm} % just in case for shifting the figure slightly down 
\includegraphics[width=7cm]{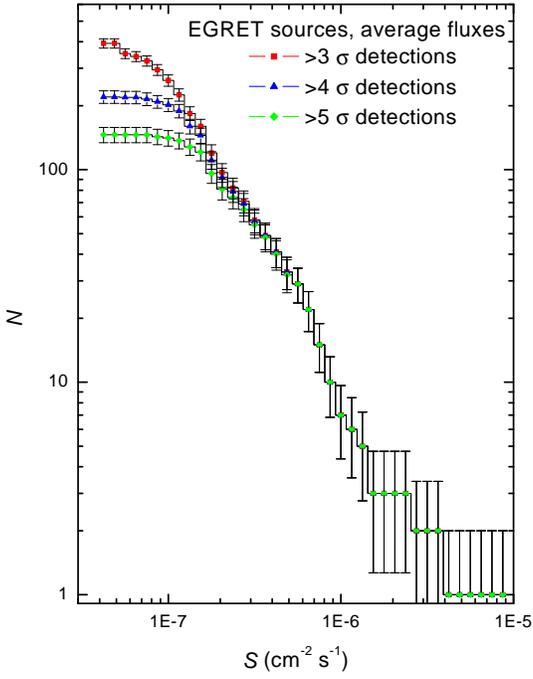}  
\caption{LogN--logS distributions of gamma-ray sources at different 
detection significance thresholds.}
\end{figure}
\begin{figure}[t] 
\vspace*{2.0mm} % just in case for shifting the figure slightly down 
\includegraphics[width=7cm]{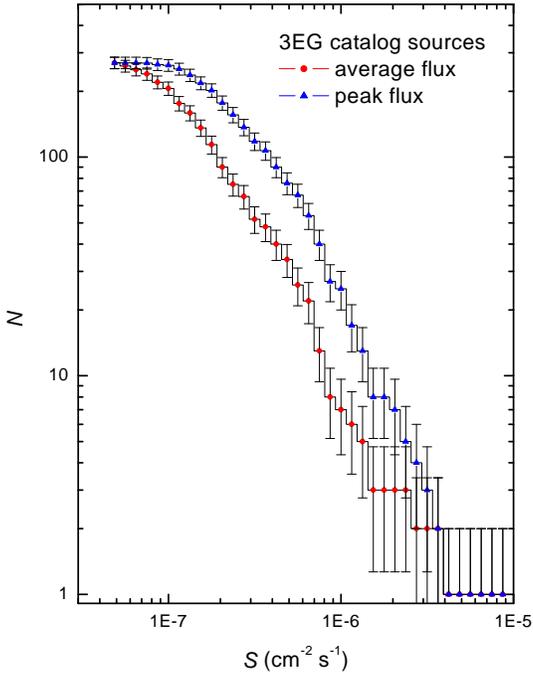}  
\caption{LogN--logS distributions comparing 3EG catalog sources regarding peak
flux and average flux values, respectively.}
\end{figure}
\begin{figure}[t] 
\vspace*{2.0mm} % just in case for shifting the figure slightly down 
\includegraphics[width=7cm]{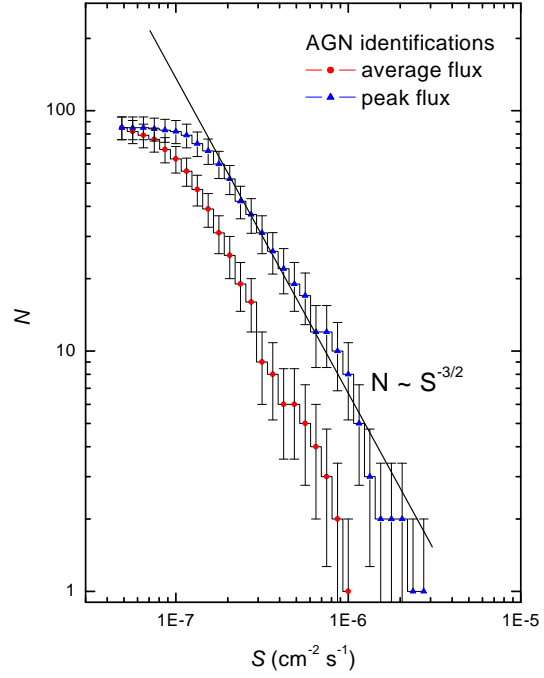}  
\caption{LogN--logS distributions comparing identified AGN regarding peak
flux and average flux values, respectively.}
\end{figure}
\begin{figure}[t] 
\vspace*{2.0mm} % just in case for shifting the figure slightly down 
\includegraphics[width=7cm]{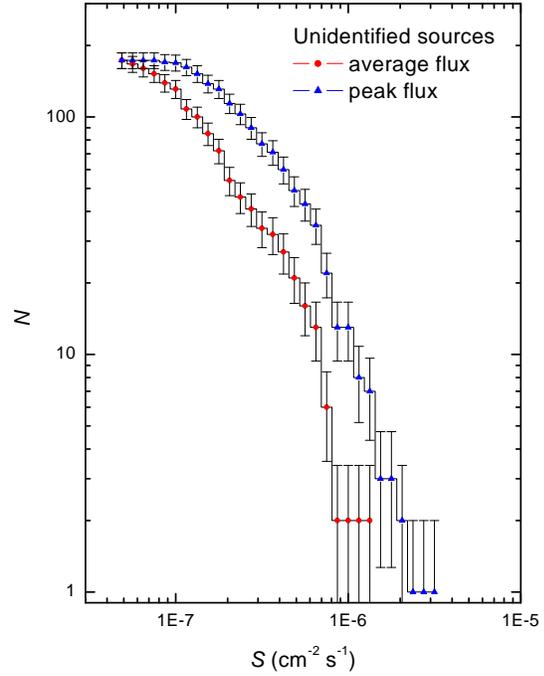}  
\caption{LogN--logS distributions comparing unidentified gamma-ray sources from
the 3EG catalog regarding peak flux and average flux values, respectively.}
\end{figure}
\begin{figure}[t] 
\vspace*{2.0mm} % just in case for shifting the figure slightly down 
\includegraphics[width=7cm]{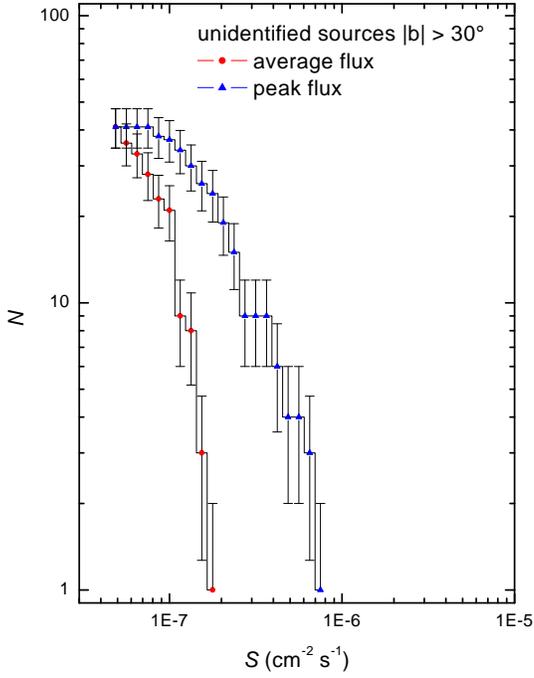}  
\caption{LogN--logS distributions comparing high-galactic latitude unidentified sources 
regarding peak flux and average flux values, respectively.}
\end{figure}
\begin{figure}[t] 
\vspace*{2.0mm} % just in case for shifting the figure slightly down 
\includegraphics[width=7cm]{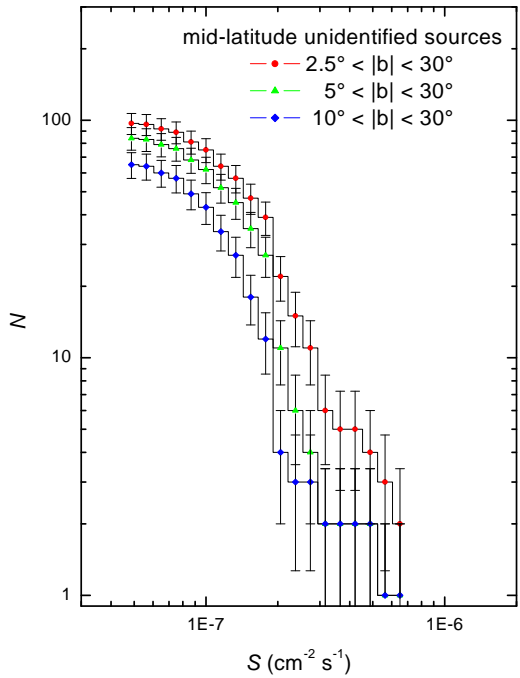}  
\caption{LogN--logS distributions comparing different latitudinal cuts describing 
mid-latitude unidentified sources ("steady" sources, "persistent" sources).}
\end{figure}
\begin{figure}[t] 
\vspace*{2.0mm} % just in case for shifting the figure slightly down 
\includegraphics[width=7cm]{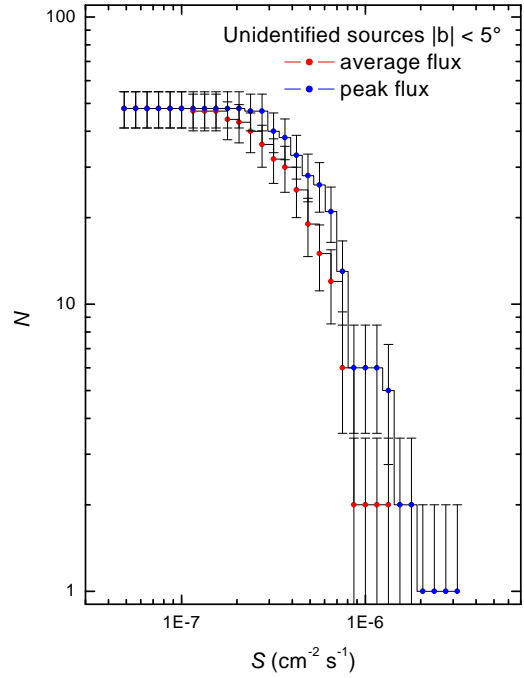}  
\caption{LogN--logS distributions comparing unidentified EGRET sources close to the
Galactic Plane regarding peak flux and average flux values, respectively.}
\end{figure}
\begin{figure}[t] 
\vspace*{2.0mm} % just in case for shifting the figure slightly down 
\includegraphics[width=7cm]{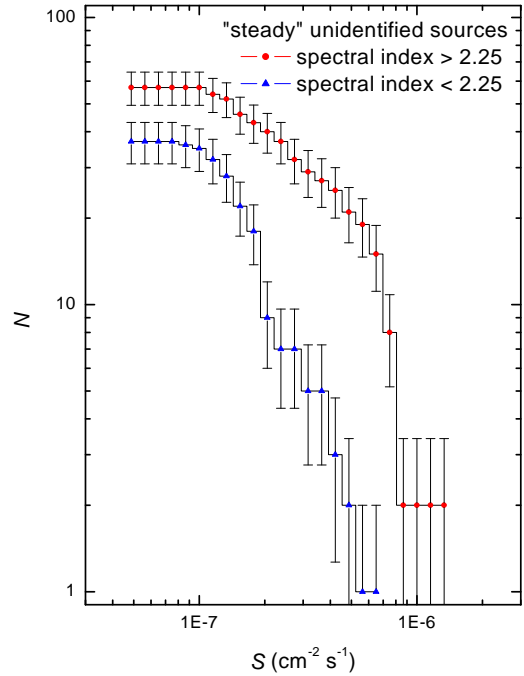}  
\caption{LogN--logS distributions comparing unidentified EGERT sources 
regarding different spectral hardness.}
\end{figure}

\end{document}